\documentclass[hyper,11pt,letterpaper]{JHEP3}
\usepackage[dvips]{epsfig}
\usepackage{amsmath,amssymb,epsf,amsfonts}
\usepackage{graphicx}
\usepackage{dcolumn}
\usepackage{bm}
\usepackage{subfig}
\usepackage{graphicx}

\usepackage[english]{babel}
\usepackage{amssymb}
\usepackage{amsfonts}
\usepackage{amsmath}
\usepackage{graphicx}
\usepackage{epsfig}
\usepackage{cite}
\newcommand{\be}{\begin{equation}} \newcommand{\ee}{\end{equation}}
\newcommand{\bea}{\begin{eqnarray}} \newcommand{\eea}{\end{eqnarray}}
\newcommand{\beann}{\begin{eqnarray*}}  \newcommand{\eeann}{\end{eqnarray*}}
\newcommand{\bfig}{\begin{figure}} \newcommand{\efig}{\end{figure}}
\newcommand{\ba}{\begin{array}} \newcommand{\ea}{\end{array}}
\newcommand{\bcen}{\begin{center}} \newcommand{\ecen}{\end{center}}
\newcommand{\btab}{\begin{tabular}} \newcommand{\etab}{\end{tabular}}

\newtheorem{Proposition}{Proposition}[section]

\newtheorem{Theorem}{Theorem}[section]
\newtheorem{Lemma}{Lemma}[section]
\newtheorem{Corrolary}{Corrolary}[section]

\newcommand{\bp}{\begin{Proposition}}   \newcommand{\ep}{\end{Proposition}}
\newcommand{\bt}{\begin{Theorem}}   \newcommand{\et}{\end{Theorem}}
\newcommand{\bl}{\begin{Lemma}}     \newcommand{\el}{\end{Lemma}}
\newcommand{\bc}{\begin{Corrolary}} \newcommand{\ec}{\end{Corrolary}}
\title{
Novel Solutions of Finite-Density D3/D5 Probe Brane System
and Their Implications for Stability
}
\author{
Han-Chih Chang \,
and
Andreas Karch
\\
Department of Physics, University of Washington, Seattle, WA 98195-1560\\
E-mail: \email{hanchih@uw.edu, akarch@uw.edu}
}

\abstract{
In this article we present a novel set of solutions of the probe brane system consisting of
$N_{f}$-D5 probe branes embedded in the near-horizon geometry generated by $N_{c}$-D3 branes,
with the D5 worldvolume $U(1)$ gauge fields turned on. Our system is holographically dual to
a supersymmetric defect field theory at finite density in non-trivial vacua.
We find that a large class of vacua turns out to satisfy a no-force condition,
even with supersymmetry explicitly broken by the finite density;
our solutions include configuration in which charge separates from the horizon and is instead
carried by probe branes outside the horizon.
The free energy is lowered in this process. Whether this corresponds to a genuine instability of
the finite-density probe brane system remains to be seen.
}

\begin{document}
\maketitle

\section{Introduction} 

Gauge/Gravity duality
\cite{Maldacena:1997re,Gubser:1998bc,Witten:1998qj,Aharony:1999ti}
has been extensively deployed as a tool to understand
the universal features shared among strongly coupled gauge theories,
with various models proposed for
quark-gluon plasma
\cite{Policastro:2001yc,CasalderreySolana:2011us},
high-$T_{c}$ superconductors
\cite{Herzog:2007ij,Hartnoll:2008vx,Hartnoll:2008kx},
unitary Fermi gas
\cite{Son:2008ye,Balasubramanian:2008dm,Schafer:2009dj}, etc.

Originally realized through Maldacena's identification \cite{Maldacena:1997re} between
$\mathcal{N}=4$ supersymmetric Yang-Mills field theory and
type IIB string theory on an AdS$_5\times$S$^5$ background,
through the dual descriptions of strong and weak coupling limits of
the dynamics of D3 branes,
the correspondence has been vigorously extended,
including conformal field theories of various internal symmetry
dual to supergravity on AdS$_{5}$ backgrounds with
corresponding internal  manifolds.
Most relevant to this article, the correspondence also has been extended to
include matter fields
\cite{Karch:2000gx,Karch:2002sh},
by inserting $N_f$ Dq probe (flavor) branes intersecting the  $N_c$  D3 background (color) branes.
Under the probe brane limit, $N_f \ll N_c$,
the gravitational back-reaction of the Dq probe branes on the background metric can be neglected.
In the weak coupling description of the Dp/Dq system,
the probe brane worldvolume SU($N_f$) gauge fields
decouple as long as $q>p$,
generating a global (rigid) flavor symmetry.
%
This model hence admits a dual interpretation of gauge theories with
fundamental field content. The $N_f \ll N_c$ limit corresponds to the quenched approximation.

Of special interest to us is the probe brane system consisting of intersecting D3/D5 branes \cite{Karch:2000gx,DeWolfe:2001pq}.
The dual gauge theory picture is the flavoring of the bulk $\mathcal{N}=4$ SYM theory
by the introduction of fundamental representation matter confined on a codimension-one defect. The localized matter consists of
$N_f$ $\mathcal{N}=2$ supersymmetry-preserving hypermultiplets.
In the weakly coupled brane system,
these extra fundamental degrees of freedom are provided by the D3-D5 open strings.
In the gravity side, only the gauge invariant operators including at least two flavor fields are visible;
their spectrum is encoded in the fluctuations of the 5-5 fields in the bulk.
The preserved supersymmetry manifests itself through a no-force condition
(this can also be shown explicitly using the $\kappa$ symmetry).

The Higgs phase of the probe brane system is also well-known
\cite{DeWolfe:2001pq,Erdmenger:2002ex,Arean:2006vg,Gaiotto:2008sa}.
By turning on the probe brane worldvolume field strength,
the background Ramond-Ramond 4-form potential
is coupled to the probe brane action through the Wess-Zumino term.
This is equivalent to separating some of the color D3 branes from the stack of $N_c$ D3 branes
hiding behind the horizon and dissolving them into fluxes inside
the flavor branes carrying corresponding D3-brane charges.
This has been realized  by introducing probe brane worldvolume magnetic flux in the D3/D5 system,
and instanton fields in the D3/D7 system.
In the gauge theory picture, this splitting of color branes partially breaks the color gauge symmetry,
and induces the vacuum expectation values of the quark content bounded at the defect.


In this paper we are mostly interested in brane systems corresponding to the field theory at finite baryon number chemical potential.
In the holographic investigation of the finite-density quantum field theories,
one constructs the dual gravity solution by turning on the probe brane worldvolume electric flux
to generate the non-vanishing time component of the gauge potential at the boundary.
Latter is dual to non-vanishing chemical potential in the field theory through the AdS/CFT dictionary.
Since the resulting field theory system retains finite entropy even at zero temperature,
there is a macroscopic degeneracy associated with the vacuum state.
While this degeneracy may well be just an artifact of the large $N_c$ and large 't Hooft coupling limit
underlying our construction, it may also indicate that we are expanding around the wrong vacuum and hence a potential instability under fluctuations.
Perturbative stability of the D3/D7 probe brane system has been investigated by Ammon, et al. \cite{Ammon:2011hz},
where they found that in spite of all the arguments favoring instability,
the D3/D7 probe brane system remains stable within the realm of perturbative fluctuations.
To investigate the non-perturbative fluctuations, however,
one will need to obtain the information about other possible configurations smoothly connected to the
fiducial vacuum state dual to the gravity solution.
One potential instability is towards disintegration of the central stack of D3 branes;
probe D3 branes could separate from the bulk horizon and carry some fraction of the baryon number with them.
In the D3/D7 system these probe D3 branes should dissolve themselves as instantons inside the flavor D7 brane,
moving the system into the Higgs phase \cite{Erdmenger:2002ex}.
In order to see whether the system is stable, meta-stable or unstable towards disintegration of the central D3 brane stack,
we are hence required to study the system on its Higgs branch at finite density.
For the D3/D7 system, progress in this direction is hampered by the fact that one is dealing with instanton configurations.
In the D3/D5 system, the Higgs branch simply corresponds to magnetic flux and is more tractable.
Nevertheless, one is being faced with having to solve complicated non-linear partial differential equation derived
from the Born-Infeld action.


In this paper,
we will present a novel set of solutions for the  D3/D5 probe brane system
with non-trivial worldvolume field strength. We construct an ansatz for the worldvolume fields
which reduces the full set of equations of motion into a single equation,
the celebrated Born-Infeld equation,
constraining only the probe brane worldvolume electric field strength.
The corresponding vacuum state turns out to obey a no-force condition,
hence providing us a sector of connected vacua around the fiducial finite-density vacuum
used in the AdS/CFT correspondence dictionary.
We show that the finite-density theory can lower its free energy by moving out on the Higgs branch. Whether this corresponds to a genuine instability remains to be seen.
We will review
the settings of the D3/D5 probe brane system in Section~\ref{D3D5setup},
present the ansatz we find in Section~\ref{ansatz},
and discuss the stability of the finite-density configuration in Section~\ref{discussinstablility}.


\section{Review of The D3/D5 Probe Brane System \label{D3D5setup}} 
In this section we will briefly review the dual gravity construction of the defect conformal field theory.
The complete dictionary can be found in \cite{DeWolfe:2001pq}.
The intersecting D3/D5 probe brane is given by the following probe configuration in the
flat 10-dimensional Minkowski spacetime:
\begin{center}
\begin{tabular}{|c||c|c|c|c|c|c|c|c|c|c|}
\hline
	& $x^0$ & $x^1$ & $x^2$ & $x^3$ & $x^4$ & $x^5$ & $x^6$ & $x^7$ & $x^8$ & $x^9$\\
\hline
D3	&x&x&x&x&o&o&o&o&o&o\\
\hline
D5	&x&x&x&o&x&x&x&o&o&o\\
\hline
\end{tabular}
\end{center}
\noindent
In the supergravity limit, the near-horizon metric generated by the $N_c$ D3 color brane is given by:
\begin{equation}
ds^2
\equiv
g_{\mu \nu} dx^{\mu} dx^{\nu}   
= \frac{1}{{r_6}^2} (-dt^2 +d\vec{x}^2)+\frac{1}{{r_6}^2}(d{r_6}^2+{r_6}^2 d\Omega_5^2),
\end{equation}
where $r_6$, the AdS radial direction, is given by:
\begin{equation}
r_6=\sqrt{(x^4)^2+(x^5)^2+(x^6)^2+(x^7)^2+(x^8)^2+(x^9)^2}.
\end{equation}
There is no dilaton associated with the supergravity background generated by the D3 color branes;
however,
the background contains the Ramond-Ramond 5-form field strength,
with the Ramond-Ramond 4-form potential given by:
\begin{equation}
C^{(4)}=\frac{r^4}{R^4} dx^0\wedge  dx^1\wedge  dx^2\wedge  dx^3.
\end{equation}
We are considering the probe brane limit, $N_{f} \ll N_{c} $,
equivalent to ignoring the back-reaction of the D5 probe branes upon the background geometry.
We also turn on the probe brane worldvolume field strength associated with the U(1) factor
of the U($N_{f}$) gauge group on $N_{f}$  coincident D5 probe brane, anticipating the
application as a dual description of a defect field theory with non-vanishing U(1)$_B$ charges.
In the probe limit,
the D5 probe branes simply minimize their worldvolume action in a fixed background.
The action is the sum of the corresponding Dirac-Born-Infeld action describing the geometric embedding within the throat geometry
and the Wess-Zumino term associated with the background Ramond-Ramond 4-form potential:
\begin{eqnarray}
S_{D5}^{DBI}	&=&  - N_{f} T_{D5} \int d^6\xi \sqrt{-   \det\left( P[g]+\tilde{F} \right)}, \\
S_{D5}^{WZ}	&=&    N_{f} T_{D5} \int P[C]\wedge \tilde{F}.
\end{eqnarray}

Using static gauge, $(\xi^0,\xi^1,\xi^2,\xi^3,\xi^4,\xi^5) \equiv (x^0,x^1,x^2,x^4,x^5,x^6)$,
the bulk field content includes:
the  embedding function for the boundary coordinate,
$ x^3(\xi^i) \equiv \phi(\xi^i) $;
the embedding functions for the internal bulk coordinate,
$x^7(\xi^i) \equiv  \eta(\xi^i) $  and $x^{8,9}(\xi^i)$;
the D5 world volume field strengths, $F= B + E \wedge dt$.
%
%
To maintain the translational symmetry of the boundary theory,
all the dependence 
on the boundary coordinates will be eliminated,
hence the bulk fields only depend on the internal space coordinates, $\xi^{3,4,5}$.
We will  be looking for the dual theory with supersymmetry preserving massive hypermultiplet. The mass
corresponds to non-zero separation between D5 and D3 branes.
In the weakly coupled brane setup, the open strings spanning from D5 to the D3 branes give rise to the featured massive hypermultiplets. A simple supersymmetric solution to the equations of motion is obtained by setting
$x_7$, $x_8$ and $x_9$ to be constant \cite{Karch:2002sh}.
We will utilize the $SO(3)$ symmetry of the internal symmetry to rotate the solution so that $x^7=m$, $x^8=x^9=0$.
$m$ is a free parameter; effectively we see that our system realizes a no-force condition. This is a manifestation of the supersymmetry preserved by this setup.

\section{The Novel Solutions \label{ansatz}} 

In this section we will present the main result of this paper,
the novel solutions for the probe brane system
with the presence of the probe brane worldvolume electric and magnetic field strengths.
In the following,
given that the bulk field content only depends on the 3-dimensional internal space,
we will abuse notation and denote the formulae in vector notation for differential operators
associated the internal coordinates.

\subsection{Solutions with Only Non-Trivial Worldvolume Magnetic Field Strength Present\label{sec:pureB}}

We first investigate the situation with only the world-volume magnetic fields turned on,
upon which the Lagrangian densities of the D5 probe brane action can be reduced into the following forms:
\begin{eqnarray}
\mathcal{L}_{D5}
&=&
\mathcal{L}^{DBI}_{D5}+
\mathcal{L}^{WZ}_{D5},
\\
\mathcal{L}^{DBI}_{D5}
&=&
-
\sqrt{
1 + {r_6}^4 	(
			|\vec{B}|^2 + |\nabla \phi|^2
		)
+ {r_6}^8 (\vec{B}\cdot \nabla \phi)^2
},
\label{lagrangian:DBIpureB}
\\
\mathcal{L}^{WZ}_{D5}
&=&
{r_6}^4 \vec{B} \cdot \nabla \phi.
\label{lagrangian:WZ}
\end{eqnarray}

This complicated system of partial differential equations however
admits one particular set of solutions:
guided by BPS techniques, where the second-order operator is decomposed into product of first-order operators,
we notice that the Lagrangian density only depends on the derivatives of boundary embedding function $\phi(\xi^{i})$,
hence the functional derivative of the action with respect to $\phi$ field gives:
\begin{equation}
\partial_{i} \left( \frac{\partial \mathcal{L}}{\partial \phi_{,i}}\right) =0.
\label{ansatzeqnold}
\end{equation}
A simple ansatz for solving this equation is to require that the variation of the action with respect to the gradient of the scalar field vanish identically (and not just its divergence).
\begin{equation}
\frac{\partial \mathcal{L}}{\partial \phi_{,i}}=0 .
\label{ansatzeqn}
\end{equation}
This ansatz can be simplified into the following relation between the magnetic fields and the boundary embedding function:
\begin{eqnarray}
0
= \frac{\partial \mathcal{L}}{\partial \phi_{,i}}
&=&
\frac
{- {r_6}^4 \phi_{,i} - {r_6}^8 (\vec{B}\cdot\nabla\phi) B_{i}
}
{
\sqrt{
1 + {r_6}^4 	(
			|\vec{B}|^2 + |\nabla \phi|^2
		)
+ {r_6}^8 (\vec{B}\cdot \nabla \phi)^2
}
}
+
{r_6}^4 B_{i}
\nonumber
\end{eqnarray}
\begin{eqnarray}
\Rightarrow
\vec{B} =\nabla \phi.
\label{solpureB}
\end{eqnarray}
With this ansatz explicitly spelled out, it now becomes straightforward to check that
the rest of equations of motion are solved:
due to the symmetry between $B_{i}$ and $\phi_{,i}$ in the Lagrangian density,
it is easy to see that the equations of motion for gauge potentials are satisfied:
\begin{eqnarray}
0=
\partial_j \left( \frac{\partial \mathcal{L}}{\partial A_{i,j}} \right)
=
\partial_j \left(
\frac{\partial  B_k}{\partial  A_{i,j}}
\frac{\partial \mathcal{L}}{\partial  B_{k}}
\right);
\label{eq:eomB}
\end{eqnarray}
The equation of motion for $\eta$ field (that is the $x^{7}$ coordinate) is also solved for any constant $m\equiv\eta_0$ assignment:
\begin{eqnarray}
0=
\frac{\partial \mathcal{L}}{\partial \eta}
&=&
\frac{\partial {r_6}^4}{\partial \eta}
\frac{\partial \mathcal{L}}{\partial {r_6}^4}
\nonumber
\\
&=&
\frac{\partial {r_6}^4}{\partial \eta}
\left(
\frac{1}{2}
\frac{
- (|\vec{B}|^2 + |\nabla \phi|^2)
-2 {r_6}^4 (\vec{B}\cdot \nabla \phi)^2
}{
\sqrt{
1 + {r_6}^4 	(
			|\vec{B}|^2 + |\nabla \phi|^2
		)
+ {r_6}^8 (\vec{B}\cdot \nabla \phi)^2
}
}
+
\vec{B}\cdot\nabla{\phi}
\right).
\label{eq:eometa}
\end{eqnarray}
It is easy to check that with our solution for $\phi$ and $\vec{B}$ the expression in parentheses vanishes identically.

Notice that the ansatz Eq.(\ref{solpureB}) completes the square of
the DBI part of the Lagrangian density Eq.(\ref{lagrangian:DBIpureB}),
which cancels all nontrival field dependence from the Wess-Zumino part Eq.(\ref{lagrangian:WZ}).
This leads to a constant on-shell Lagragian density irrespective of any detail of the solutions within this sector,
not even the mass scale $m$ associated with the transverse splitting
between D3 and D5 branes. This is of course a signal of the preserved supersymmetry. These solutions
correspond to the supersymmetry preserving Higgs branch of the dual defect field theory.

To understand the geometry of these solutions, note that we can turn on any arbitrary boundary embedding function,
$\phi=\sum_{i} \frac{q_i}{|r_3-r_i|}$
( $r_3 \equiv\sqrt{\xi^3+\xi^4+\xi^5}$), while preserving the no-force condition.
The corresponding $\frac{1}{|r_3 - r_i|}$ spikes in the scalar field at the positions $r_i$ can be interpreted
as D3-branes ending on the D5 probe brane. The endpoint of the D3 brane sources the magnetic worldvolume field which
(through the Wess-Zumino action  Eq.(\ref{lagrangian:WZ})) ensures that the D5 brane carries D3 brane charge and
so we don't violate D3 charge conservation.
Our bulk solution describes hence an arbitrary number of half D3 branes ending on either side of the
D5 branes at arbitrary positions in the 3d internal space.
We can label the general solution by the positions $r_i$ and the numbers $q_i$ of D3 branes ending at at given $r_i$.
The signs of the $q_i$ indicate in which side of the D5 a given half-D3 ends.
Schematically such a supersymmetric Higgs branch configuration is displayed in panel (a) of figure~\ref{fig:animals}.

The special case of a single stack of half-D3 branes ending at the origin
$\nabla \cdot B =\nabla^2{\phi} = Q\delta^{3}(r)$
while setting $m=0$
has been analyzed previously \cite{Karch:2000gx,Arean:2006vg,Myers:2008me}.
Using the spherical symmetry of the system, this can be recast as an embedding wrapping a constant $S^2$ inside
$S^5$ with $Q$ units of worldvolume magnetic flux piercing the $S^2$.
The embedding coordinate $x^3$ (described by the field $\phi$) now only depends on the radial coordinate.
As $\phi$ only appears in the equations derivatively, one can simply integrate the resulting ordinary differential equation; the result is in complete agreement with our general solution.
In this special case, one can even find the corresponding finite-temperature solution \cite{Myers:2008me}.

The single magnetic charge at the origin with non-zero mass is also a very interesting special solution,
first discussed in \cite{Arean:2006vg}.
In the weakly coupled brane picture, we have a single half-D3 brane ending on the stack of $N_f$ D5 branes,
but the whole D5/half-D3 merged brane configuration is separated from the stack of $N_c$ color D3 branes
by a distance proportional to the mass.
In the holographic dual description, we have a probe brane that is not touching the horizon.
The D5 brane turns into a single spike running parallel to the horizon off to infinite $x_3$.
Waves incoming from the boundary can be supported with any frequency, so the system has no gap.
This seems to be the only known example of a probe brane that supports massless fluctuations but doesn't touch the horizon.
The fluctuations originally localized on the D5 can run off to spatial infinity in
the transverse $x_3$ direction.


\subsection{Non-Trivial Worldvolume Electric and Magnetic Field Strength}
Introducing the electric field strength $\vec{E}$ on the D5 probe brane worldvolume will not
modify the coupling of the background Ramond-Ramond 4-form potential from Wess-Zumino part of the action Eq.(\ref{lagrangian:WZ}),
but the DBI part becomes more complicated:
\begin{eqnarray}
\mathcal{L}^{DBI}_{D5}
=
-\sqrt{
1-|\vec{E}|^2
+ {r_6}^4 	(
			|\vec{B}|^2 + |\nabla \phi|^2 - (\vec{B} \cdot \vec{E})^2 - |\vec{E}\times\nabla\phi|^2
		)
+ {r_6}^8 (\vec{B}\cdot \nabla \phi)^2
}.
\label{lagrangian:DBIwithE}
\end{eqnarray}
We will employ the similar reduction method as before, that is we demand that the variation of the action
with respect to the gradient of the scalar field vanish identically. Solving Eq.(\ref{ansatzeqn}) we find:
\begin{eqnarray}
\vec{B}&
=
\sqrt{1-|\vec{E}|^2} \nabla\phi +
\frac{\vec{E}\cdot\nabla\phi}{\sqrt{1-|\vec{E}|^2}}\vec{E}.
\label{ansatz:withE}
\end{eqnarray}
With this relation, a tedious but straightforward algebraic exercise reveals that
the full set of equations of motion is reduced into a single constraint.
More explicitly,
one finds that this ansatz alone not just solves the equations of motion for the $\phi$ field. In addition, once more the equations of motion for
the $\eta$ field (the analog of Eq.(\ref{eq:eometa}))
and the internal-spacial components of the gauge potential $A_i, \forall i\in \{3,4,5\}$
(the analog of Eq.(\ref{eq:eomB})) are automatically fulfilled.
But the equation of motion for the time-component of the gauge potential $A_t$ demands an additional condition to hold,
which turns out to be the celebrated Born-Infeld equation on the electric field strength:
\begin{eqnarray}
\nabla \cdot \left(\frac{\vec{E}}{\sqrt{1-|\vec{E}|^2}}\right) =0.
\quad
\label{equDBI}
\end{eqnarray}
Therefore, the solution can be reformulated as the problem of finding the Born-Infeld electrostatic solutions.

The on-shell Lagrangian density,
the sum of Eq.(\ref{lagrangian:DBIwithE}) and Eq.(\ref{lagrangian:WZ}),
can be further simplified using the relation Eq.(\ref{ansatz:withE}):
\begin{equation}
\mathcal{L}^{On Shell}=\sqrt{1-|E|^2}.
\end{equation}
Notice that the on-shell Lagrangian density only depends on the  electric field strength $\vec{E}$,
but not on the boundary embedding function $\phi$. The simplest analytic solution to this equation of motion
is that of a single point charge located at the origin,
\begin{eqnarray}
\label{bion}
\vec{E}=\frac{Q}{\sqrt{{r_3}^4+Q^4}}\hat{r}_3,
\end{eqnarray}
with an electric field decaying with the usual $1/r$ at large distances but reaching a finite value at small $r$.
For the special case that no magnetic fields (and hence no $\phi$ field) are turned on, this finite-density solution
matches the analytic solution previously studied e.g. in \cite{Karch:2007br,Karch:2008fa}.
Supersymmetry is explicitly broken in these solutions.
On the field theory side, this is due to the presence of a finite chemical potential.
The free energy is negative. What our solution shows is that even in the presence of this finite density,
we still have the full moduli space of half-D3 branes ending at arbitrary positions $r_i$
we found in the zero density supersymmetric vacuum.
A typical configuration of this type is displayed in panel (b) of figure~\ref{fig:animals}.
Irrespective of the position of the D3-brane spikes,
the free energy of the system is always given by the expressions found in \cite{Karch:2007br},
as long as the electric field is given by the simple BIon like solution Eq.(\ref{bion}).

Note that, as in the last subsection, further progress can be made in the spherically symmetric case, that is when $\phi$ and $\vec{B}$ are due to a single source at the origin. In this case the magnetic field once more is simply given by a constant flux through the internal $S^2$.
One obtains coupled ordinary differential equations for $A_t(r)$ and $\phi(r)$ which can be integrated as both fields only appear derivatively in the action.
One can easily reproduce the interesting feature that the free energy, at zero temperature,
is complete independent of the magnetic flux through the sphere and only depends on the baryon number chemical potential,
that is the electric flux. In this special case one can also find the finite-temperature embeddings
\footnote{These solutions
have been worked out in collaboration with Andy O'Bannon and will potentially appear elsewhere.}.

This intriguing presence of supersymmetric-broken moduli space demands further investigation.
What is perhaps even more interesting is to study other solutions of the Born-Infeld equations corresponding to multiple separated electric charges. These multiple-BIon solutions
and
their implication for stability will be the main topic for the next section.



\section{DBI Solutions Revisited and the Stringy Interpretations \label{discussinstablility}} 
The ansatz we constructed in the last section reduces the full set of nonlinear coupled equations of motion into
the single Born-Infeld equation, Eq.(\ref{equDBI}), allowing us to explore
the dynamics of the probe brane world volume gauge field.
It is well known that no non-trivial global regular solutions of Eq.(\ref{equDBI}) do exist,
see e.g. \cite{Gibbons:1997xz}.
To show this, one first observes that solutions of Born-Infeld electrostatic can be shown, by the level curve technique, to be the same as maximal hypersurfaces in Minkowski spacetime. For latter
a theorem exist (the Minkowski version of the original Bernstein theorem) saying that the regular solutions can only be trivial.
Therefore, to introduce the non-trivial electric field strength,
we will need to modify the right hand side of the  Eq.(\ref{equDBI}) by explicitly including source terms.
We have not introduced any charged fields in the bulk theory. So all such sources have to either sit behind the horizon or at spatial infinity, that is at infinite $x_3$ in the AdS spacetime.

The simplest non-trivial solution is given by a single BIon, Eq.(\ref{bion}). We needed to introduce a single point charge $Q$ in order to support this solution. This charge is located at $r=0$, which corresponds to the horizon in the AdS geometry. As in most holographic finite-density studies, the electric field of the single-BIon is sourced by a charged horizon. Standard lore has it that this corresponds to charges
carried by fractionalized charge carriers in the field theory \cite{Hartnoll:2011pp}.
The corresponding gauge potential $A_{t}$ approaches an asymptotic constant at the boundary,
dual to the chemical density of the associated vacuum state in the
dual gauge theory description of AdS/CFT dictionary.

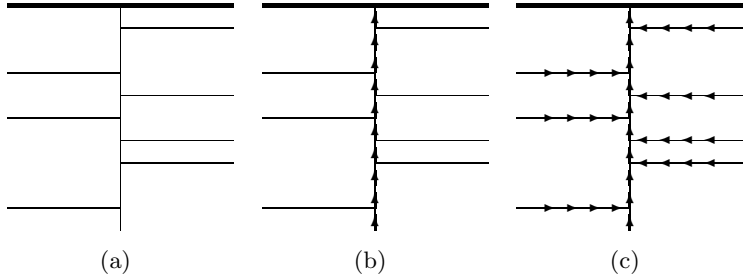
\begin{figure}
\centering
\subfloat[]{\label{fig:gull}
\setlength{\unitlength}{3mm}
\begin{picture}(10, 12)(-5,-10)
\linethickness{0.5mm}
  \put(-5,0){\line(1,0){10}}
\linethickness{0.05mm}
  \put(0,0){\line(0,-1){10}}
\linethickness{0.05mm}
  \put(0,-3){\line(-1,0){5}}
  \put(0,-5){\line(-1,0){5}}
  \put(0,-9){\line(-1,0){5}}
  \put(0,-1){\line(1,0){5}}
  \put(0,-4){\line(1,0){5}}
  \put(0,-6){\line(1,0){5}}
  \put(0,-7){\line(1,0){5}}
\end{picture}
}
\subfloat[]{\label{fig:tiger}
\setlength{\unitlength}{3mm}
\begin{picture}(10, 12)(-5,-10)
\linethickness{0.5mm}
  \put(-5,0){\line(1,0){10}}
\linethickness{0.05mm}
  \put(0,0){\line(0,-1){10}}
\linethickness{0.3mm}
  \multiput(0,-10)(0,1){10}{\vector(0,1){0.7}}
\linethickness{0.05mm}
  \put(0,-3){\line(-1,0){5}}
  \put(0,-5){\line(-1,0){5}}
  \put(0,-9){\line(-1,0){5}}
  \put(0,-1){\line(1,0){5}}
  \put(0,-4){\line(1,0){5}}
  \put(0,-6){\line(1,0){5}}
  \put(0,-7){\line(1,0){5}}
\end{picture}
}
\subfloat[]{\label{fig:mouse}
\setlength{\unitlength}{3mm}
\begin{picture}(10, 12)(-5,-10)
\linethickness{0.5mm}
  \put(-5,0){\line(1,0){10}}
\linethickness{0.05mm}
  \put(0,0){\line(0,-1){10}}
\linethickness{0.3mm}
  \multiput(0,-10)(0,1){10}{\vector(0,1){0.7}}
\linethickness{0.05mm}
  \put(0,-3){\line(-1,0){5}} \multiput(-1,-3)(-1,0){4}{\vector(1,0){0.7}}
  \put(0,-5){\line(-1,0){5}} \multiput(-1,-5)(-1,0){4}{\vector(1,0){0.7}}
  \put(0,-9){\line(-1,0){5}} \multiput(-1,-9)(-1,0){4}{\vector(1,0){0.7}}
  \put(0,-1){\line(1,0){5}} \multiput(1,-1)(1,0){4}{\vector(-1,0){0.7}}
  \put(0,-4){\line(1,0){5}} \multiput(1,-4)(1,0){4}{\vector(-1,0){0.7}}
  \put(0,-6){\line(1,0){5}} \multiput(1,-6)(1,0){4}{\vector(-1,0){0.7}}
  \put(0,-7){\line(1,0){5}} \multiput(1,-7)(1,0){4}{\vector(-1,0){0.7}}
\end{picture}

}
\caption{Three Different scenarios: (a) Pure Magnetic Flux; (b) A Single BIon; (c) Multiple BIons}
\label{fig:animals}
\end{figure}

Multiple-BIon solutions exist due to general existence theorem as well,
even though the explicit solutions have not been analytically proposed yet \cite{Gibbons:1997xz}.
Such multiple-BIon solutions must be supported by electric point charges located at some locations $r_a$, where $a$ runs over the point charges in the multiple-BIon solution. Naively this requires us to introduce point sources inside AdS space. Note however that as long as we chose all the $r_a$ to coincide with some of the D3-brane spikes, that is with some of the $r_i$ introduced above, we effectively pushed these charges off to infinite $x_3$. Schematically such a solution is depicted in panel (c) of figure~\ref{fig:animals}.
Note that the magnetic charge $q_i$ of the spikes is completely unrelated to the electric charge $Q_a$.
Former measures how many D3 branes end on the D5 brane, latter measures how many F1s are dissolved in it.
In the field theory side, these configurations correspond to finite baryon number on the defect sourcing a free $U(1)$ factor associated with the half-D3 brane. So such multiple-BIon solutions can easily be realized in our setup.

Furthermore,
in \cite{Gibbons:1997xz},
Gibbons argued that the single BIon, with macroscopic charge, is actually unstable against fission,
based on a scaling relation between the energy and charges of the BIons.
This fission of the macroscopic BIon may signal an instability of the finite-density
brane system with all charges behind the horizon.
Combining our result that the on-shell action of the brane system reduces to the BI action with the scaling arguments of \cite{Gibbons:1997xz} demonstrates that the free energy of a system of fixed baryon number $Q_{tot}$ is lower if
the charge is not carried by the horizon but instead is supported by D3 brane spikes. The configuration can clearly lower its free energy by disintegrating the central stack of D3 branes and letting the probe branes carry the finite density. That is, the system has lower free energy on its Higgs branch. It is not clear
to us whether this necessarily needs to be interpreted as an instability. In the process of pulling a single
probe D3 brane out of the horizon and letting it support part (or all) of the electric flux of the horizon we
change the boundary conditions on the Maxwell $U(1)$ electric field at spatial infinity. It would be very interesting
to do a  linearized analysis of fluctuations along the lines of \cite{Ammon:2011hz} for the finite-density D3/D5 system to see if this instability towards disintegration of the central D3 brane stack is visible at the
perturbative level.

\section{Future Research and Conclusion} 

We present an ansatz for the D3/D5 probe brane system with non-trivial magnetic
D5 worldvolume gauge field strength.
The ansatz turns out to
satisfy a no-force condition, signaling preserved supersymmetry. These brane configurations reproduce
the full Higgs branch moduli space of the holographically dual supersymmetric defect conformal field theory.
Within this sector, we find that probe brane system dual to the finite-density vacuum,
based on the scaling argument from in \cite{Gibbons:1997xz},
can lower its free energy by fissioning.
This potentially signals an instability whose precise nature will be left to future research.


\section*{Acknowledgements}
AK would like to thank Andy O'Bannon for collaboration during initial stages of this work and numerous helpful discussions on finite-density probe brane systems.
This work was supported in part by the US Department of Energy under contract number DE-FGO2-96ER40956.

\bibliography{ref}
\bibliographystyle{JHEP}

\end{document}